\documentclass[sigconf]{acmart}
\usepackage{subfig}
\usepackage{footmisc} 

\AtBeginDocument{%
  }

\begin{document}

\title{ Optimal Current Control Strategy for Reliable Power Electronics Converters: Frequency-Domain Approach }


\author{Amin Rezaeizadeh}
\affiliation{%
  \institution{Institute of Electric Power Systems, FHNW}
  \city{Windisch-Brugg}
  \country{Switzerland}}
\email{amin.rezaeizadeh@fhnw.ch}
\orcid{0000-0003-4625-5546}

\author{Silvia Mastellone}
\affiliation{%
  \institution{Institute of Electric Power Systems, FHNW}
  \city{Windisch-Brugg}
  \country{Switzerland}}
\email{silvia.mastellone@fhnw.ch}
\orcid{0000-0002-7731-8821}

\settopmatter{printacmref=false} 
\renewcommand\footnotetextcopyrightpermission[1]{}

\renewcommand{\shortauthors}{A. Rezaeizadeh et al.}

\begin{abstract}
Power electronics converters are  key enablers  in the global energy transition for power generation, industrial and  mobility applications; they
convert electrical power in a controlled, reliable and efficient manner.
The  semiconductor switching devices, at the core of power converters,   are the  most likely component to fail  due to the damage caused by  the current-induced  temperature cycling. 
Damage models of semiconductors have been developed and employed
to study their reliability, improve their design and  to estimate the lifetime of the converter in various power applications. However, those models can offer  more if employed in the design of   strategies to actively operate the   converter. Specifically,   properly controlling the current, and hence the temperature cycling, can effectively contribute to reducing the accumulated damage in the semiconductor and increase its reliability and lifetime.  
In this paper we propose a novel   current    control approach   that integrates   reliability requirements into the design framework, based on a frequency-domain model of the semiconductor damage. 
\end{abstract}

\keywords{Reliability control, Power converter, Thermal stress control }
\begin{teaserfigure}
  \includegraphics[width=1\textwidth]{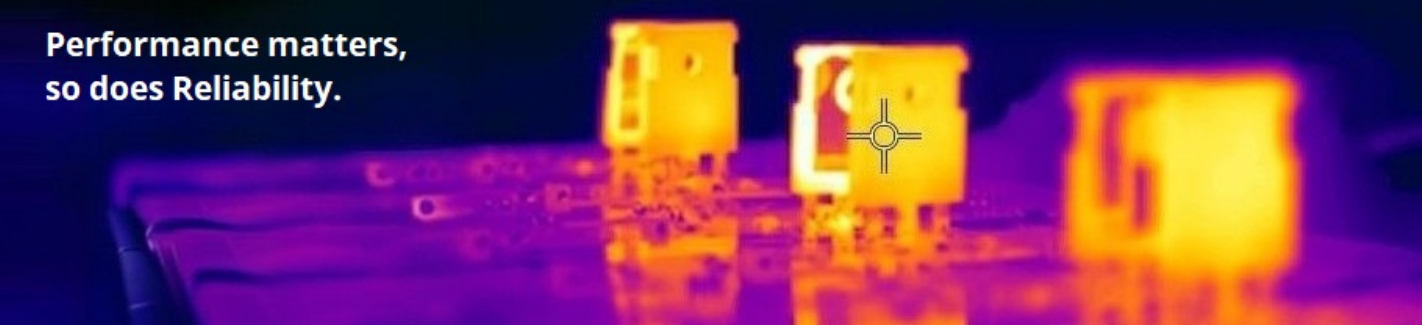}
  \caption{SiC MOSFETs experiencing thermal stress in a 2-level 3-phase power converter, captured by the FLIR thermal camera. }
  \label{fig:teaser}
\end{teaserfigure}
\acmConference{}{}{}

\maketitle
This work was supported by the Swiss National Science Foundation under NCCR Automation, grant agreement 51NF40\_180545.
\section{Introduction}
Power electronic converters are at the core of the emerging technology required to transform the energy, mobility and production industries.
They efficiently regulate the  electric power flow between energy sources and loads. In the past two decades a major research and engineering  effort has been dedicated to  develop electrical power converter systems that are efficient,  reliable, and cost-effective. The  discovery of new  materials and emerging new energy-efficient conversion technologies are supporting this  progress \cite{SiC1,PEC1,abb_3L_eff}. 

As key enabler of the ongoing electrification trends in various sectors, power electronic converters are penetrating all application sectors of our modern societies. This growth results in an increased demand for critical materials, which in turn gives rise to a range of geopolitical, economic, and societal challenges \cite{GUTSCH2022105030,Raworth2017ADF}. In order to ensure the long-term sustainability of this novel technological landscape,  it is crucial to study the  environmental impacts throughout the entire life-cycle of power converters, from design, production, operation, and finally disposal.  

Developing more sustainable power electronics converters depends on taking into account both environmental and technological indicators. It also requires the use of methods and tools to measure and analyze environmental impacts throughout the entire life cycle of the converters. The reliability measure is  one factor that can be integrated into converter sustainability studies. It enables to compute the total degradation and to  estimation of the power converter lifetime. 
In   \cite{Tang2021ReliabilityStudy}, the authors identified   power electronics converters as   the   most likely component to fail in power-trains after the battery module. Moreover,  as discussed in \cite{6532474} power converters are responsible for  13\% to 37\% of the failure and unscheduled maintenance.

At the component level, the two types of reliability-critical components are semiconductor switching devices, such as metal-oxide-semiconductor field-effect transistor (MOSFET), and  electrolytic capacitors \cite{failure1}. 
Power semiconductor devices are the  most likely component to fail in a power converter system due to their vulnerability to current-induced thermal stress  \cite{semi_survey,MIL-HDBK-217F}. This inherent vulnerability of the high-power semiconductor switches raises reliability concerns in power converters. 
Specifically,  power semiconductor devices, such as IGBTs and MOSFETs, are responsible for approximately 21\% of inverter failures \cite{failure1,failure2}. 
Thermal stress is the major cause of reduced lifetime for power switching devices, and a significant amount of research has been devoted to the electrical-thermal analysis and reliability assessment of these devices\cite{Frede1,Frede2,rel}.

The most frequent  failure in a semiconductor switch is bond wire lift-off caused by stress cracks due to thermal cycling  experienced by the device during  operation. 
As the   temperature fluctuates, the mismatch  in the thermal expansion coefficients between the aluminium wire and the silicon results in  stress at the bond interface. Consequently, the bond wires become disconnected, leading to an open-circuit failure~\cite{bondwire1,bondwire2,bondwire3}.

Any component failure  can result in unexpected system shutdowns, leading to safety risks and financial costs. Therefore, lifetime analysis and reliability-aware operation of power semiconductor devices is a critical step in power converter lifetime management and sustainability.


The performance and experienced damage   of a power converter  depends partially  on the operating conditions of semiconductor devices, which can vary depending on the application and electrical loads.
Specifically, the current profile across the semiconductors determine the level of stress and damage experienced. Regulating the current allows to control the level of stress and thus damage experienced.
Responsible system operation contributes to  reduce the system long-term damage and to increase the its lifespan. \textit{ Reliability Control} realizes this  by  employing  damage models of components and entire systems in the control design phase. 

\subsection{Related work}

 Several studies have proposed fault-tolerant control strategies to improve the safety and reliability of the power converter system in case of failure\cite{fault_tol1,fault_tol2}. Additionally,  techniques have been developed to  reduce  thermal stress and extend the converter's lifetime by actively controlling the power module's temperature\cite{cooling1,cooling2}. 
For example, in \cite{thermalControl3} proposed a varying switching frequency to reduce the thermal stress. Whereas authors in \cite{thermalControl4} use circulating current through devices to limit the cooling and thus the thermal swing.  In \cite{997541} authors studied how power devices can be operated close to their thermal limits while avoiding thermal overload. Others proposed the thermal cycle reduction by use of thermal observers \cite{801687,bakran2003aktive,Murdock,Murdock2,8540418}. An optimized thermal loading control is proposed in \cite{thermalControl1} for  the modular multilevel converter with the objective of minimizing the total power loss of the semiconductors. In our previous work we proposed optimal current and torque controllers to actively  reduce  thermal stress experienced by the semiconductors \cite{aminECC2024_mobility,aminCDC}. Other approaches design a switching strategy to control the temperature of the stressed power module and avoid overheating condition \cite{pcim,thermalControl2,thermalControl5,optimal_svm, svm2,svm3,reliability_switching}. Finally, other solutions to reduced thermal stress are based on a topology with redundant switches  \cite{redundant}.


\subsection{Paper contribution and structure}
In this work, we first study the semiconductor lifetime model in the frequency domain. Specifically, we investigate the  dominant frequency contents of the current signal that contribute the most to the semiconductors degradation. Second, we propose a current control design in the frequency domain to reduce the thermal stress while fulfilling the current reference tracking requirement.   The  frequency domain formulation allows to use the damage  model  in an online control policy, without relying  on traditional cycle counting.
We illustrate how the online reliability controller can effectively reduce the experienced damage and increase system availability and lifetime of the power converter without compromising tracking performance.

The paper is structured as follows: 
Section \ref{sec:problem} describes the system model, including the power electronics converter and  the electro-thermal characteristics of  the semiconductor modules.  Section \ref{sec:reliability}  details the semiconductor characteristics, its losses and damage models in frequency domain. Section \ref{sec:control} presents the main result, where the reliability control policy is detailed.  The controller  performance and the lifetime assessment results  are given in Section \ref{sec:simulation}.  
Finally Section \ref{sec:conc} is dedicated to concluding remarks and future research directions.

\section{System Modeling}
\label{sec:problem}
\subsection{Power converter }
\begin{figure}[h]
  \centering
  \includegraphics[width=0.6\linewidth]{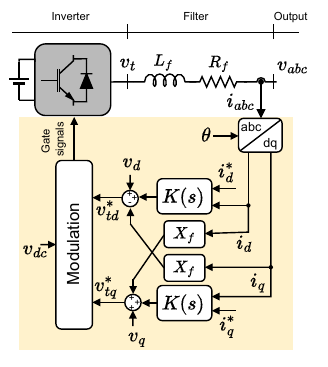}
  \caption{One-line diagram of a DC-AC inverter  showing the control system implemented in the dq-coordinate frame. \label{fig:system}}
\end{figure}

A typical DC-AC power electronics converter is illustrated in Fig. \ref{fig:system}. The three-phase inverter is composed of the switching system, which for simplicity in analysis we assume to have a 2-level topology, and the RL filter part. Ignoring the switching harmonics influence, the system dynamics in the rotating dq-frame can be described as follows:
\begin{eqnarray} \label{eq:dynamics}
    L_f \frac{di_d}{dt}&=& v_{td}-v_d -R_fi_d +L_f\omega i_q, \nonumber \\
    L_f \frac{di_q}{dt}&=& v_{tq}-v_q -R_fi_q -L_f\omega i_d,  
\end{eqnarray}
where the variables $v_d$ and $v_q$ are the dq-components of the inverter voltage, $v_{td}$ and $v_{tq}$ denote the dq-components of the inverter terminal voltage,  $i_{d}$ and $i_{q}$ are the dq-components of the inverter output current, and  $\omega$ is the angular frequency of the inverter output signals. The time-dependency is omitted for brevity. 


The three-phase current and voltage measurements are converted to the dq-coordinate frame, using the reference angle $\theta$.  The current  controllers for d and q axis are identical and considered to be a linear multi-input single-output (MISO) system. The cross term feed-forward signals are introduced to decouple the d and q dynamics (discussed in details in Sec. \ref{sec:control}). The $v^*_{td}$ and $v^*_{tq}$ are the dq-components of the desired voltage which are given as references to the modulation unit. 

\subsection{Semiconductor electro-thermal model}
\begin{figure}%
    \centering
    \subfloat[\centering ]{{\includegraphics[width=0.5\linewidth]{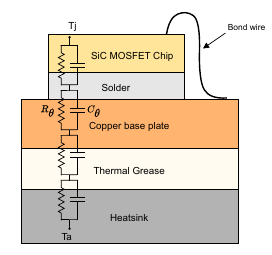}\label{fig:mosfet} }}%
    \qquad
    \subfloat[\centering ]{{\includegraphics[width=0.4\linewidth]{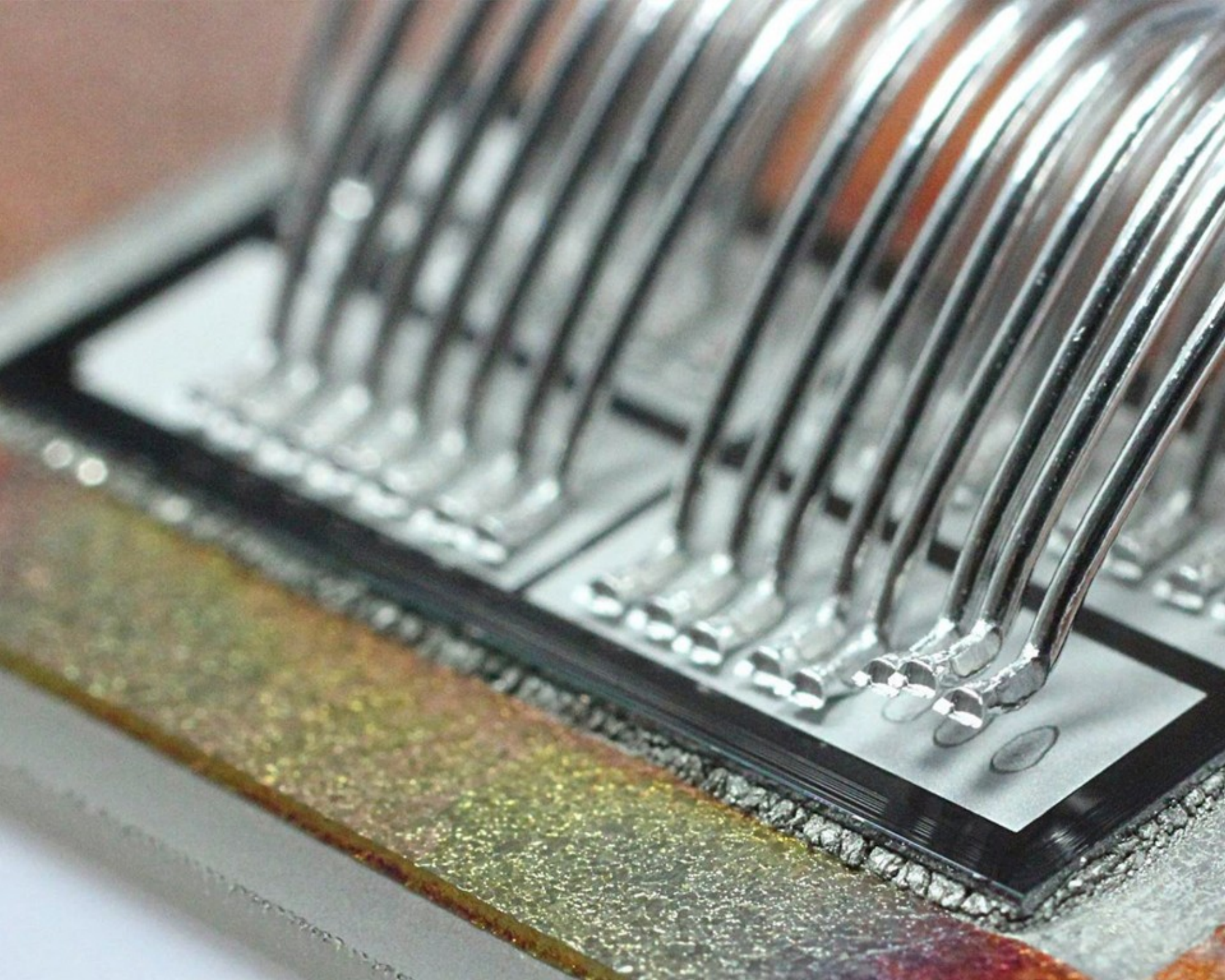} \label{fig:bondwire}}}%
    \caption{(a) A typical package of a SiC MOSFET power module. The Foster thermal model consisting of sequential RC systems represents the heat transfer through the semiconductor module. (b) The bondwire lift-off due to thermal cycling (courtesy of Fraunhofer IISB). }%
    \label{fig:example}%
\end{figure}

Figure \ref{fig:mosfet} illustrates a typical package of a SiC power module. The  heat generated by the device power loss must flow through several different layers. The transient behavior of the junction temperature, denoted by $T_j$, is governed by the time-dependent
heat diffusion equation. However, a simple approximation can be obtained using the electric circuit analog  known as the Foster thermal model \cite{mohan}.  In this model, the temperature dynamics are modeled by  a series thermal resistance  and capacitance network. The temperature difference between two consecutive layers is governed by the following linear differential equation:
 \begin{equation}
   {R_\theta} C_\theta \Delta\dot{T}(t)+\Delta T ={R_\theta} P(t),
    \label{eq:Tj}
\end{equation}
where~$R_\theta$ and~$C_\theta$ are, respectively, the thermal resistance and capacitance of the layer,~$P(t)$ denotes the power loss in  the device.

For SiC MOSFET devices, the total average power losses over the switching cycle is given by 
\begin{equation}
    P\approx r_{\mathrm{on}} I^2_{Drms} + P_{sw},
\end{equation}
where~$r_{\mathrm{on}}$ is the drain-source on-state resistance,  $I_{Drms}$ denotes the rms value of the on-state current, and $P_{sw}$ is the switching loss which for SiC MOSFETs can be negligible  compared to the conduction loss.

\section{Sustainability measure}
\label{sec:reliability} 

\subsection{Damage Model}

The number of cycles to failure can be estimated based on a lifetime model. 
In this work we use the lifetime model adopted by   Semikron  \cite{semikron}, to evaluate the semiconductor  devices reliability. This model considers only degradation of the bond wire connection to the chip, since bond wire lift-off is the dominant cause of failure in power switches (see Fig. \ref{fig:bondwire}).  The number of cycles to failure, $N_f$, of the power switches  is given by the following modified Coffin-Manson lifetime model:
\begin{equation}
\label{eq:semikron}
   N_f=A_0.A_1^\beta.\Delta T^{\alpha-\beta}.\exp{\frac{E_\mathrm{a}}{\kappa_\mathrm{B}T_j}}.\frac{C+{t_{on}}^\gamma}{C+2^\gamma}.k_{\mathrm{thick}},
\end{equation}
with
\begin{equation*}
    \beta=\exp{\frac{-(\Delta T_j -T_0)}{\lambda}},
\end{equation*}
where $T_j$ and  $\Delta T_j$ are, respectively, the average and magnitude of the  junction temperature cycle, and $t_\mathrm{on}$ is the duration of the cycle. The remaining factors are  constant parameters:  $E_\mathrm{a}$ and $\kappa_\mathrm{B}$ are, respectively, the activation energy and the Boltzmann constant, $k_{\mathrm{thick}}$ is the chip thickness factor, and $A_0$, $A_1$, $T_0$, $\lambda$, $\alpha$, $C$, $\gamma$ are empirically obtained constants given in ~\cite{semikron}. 

As the magnitude and average of the temperature cycles increase, the number of cycles to failure decreases. For example, if a SiC MOSFET is subjected to a temperature stress of $\Delta T_j=40^\circ$, it can survive approximately 300k cycles at an average junction temperature of $T_j=150^\circ$ and a duration of $t_\mathrm{on}=10s$. However, at the same temperature and cycle duration, but with a higher temperature stress of $\Delta T_j=80^\circ$, the expected
is only about 10k. It can also be seen that as the duration of the temperature cycles, $t_\mathrm{on}$, increases, the number of cycles to failure decreases slightly.

The accumulated damage $D_\mathrm{tot}$ of the semiconductor switch for one cycle of the mission profile can be determined via the Miner rule, originally proposed by Palmgren and Miner~\cite{Miner1} as 
\begin{equation}
\label{eq:total_rainflow}
D_\mathrm{tot}=\sum_{i=1}^{n} \frac{n_i(S_i)}{N_f(S_i)},   
\end{equation}
where $n_i(S_i)$ denotes the effective number of cycles at stress level  $S_i$ of the the stress history signal, and $N_f(S_i)$ denotes the available number of life-cycles depending on the material characteristics and can be calculated from (\ref{eq:semikron}). The lifetime of the power switches
can then be computed by multiplying the inverse of $D_\mathrm{tot}$ with the duration of the mission profile.

\textbf{Remark}: From the accumulated damage formula we  infer that the damage can be reduced  either by reducing the number and intensity of experienced cycles  or by increasing the number of lifetime cycles available to the semiconductors.
\begin{figure}[thpb]
      \centering
     \includegraphics[width=0.85\linewidth]{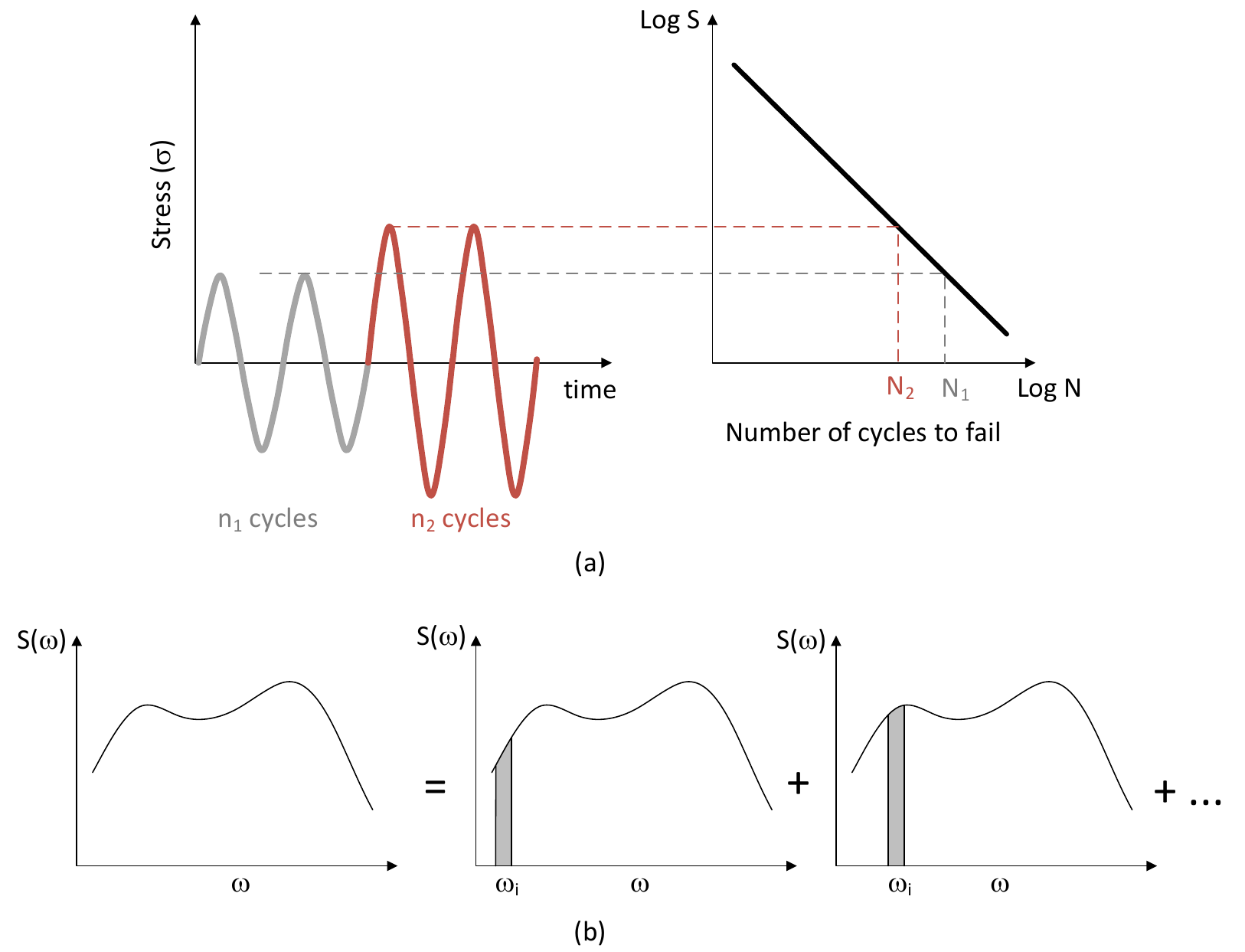}    
      \caption{Fatigue Analysis: (a) Stress versus number-of-cycles-to-failure ( the W\"{o}hler-curve or S-N curve), (b) The concept of  the single-moment method for damage estimation by  decomposing the  power spectral density (PSD) of the stress signal  into contribution of  a set of narrow-band infinitesimal spectrals.  }
      \label{fig:fatigue}
\end{figure} 
\subsection{Stress and lifetime model: S-N curve}
In material fatigue analysis, the S-N curve, introduced by Wöhler, defines  the relationship between the applied stress cycle and the number of cycles to failure for a specific material (see \ref{fig:fatigue}a). This relationship  can be described by  
the following exponential equation, commonly known as Basquin’s equation:
\begin{equation}
    N_f=c_{sn}S^{-k_{sn}},
\end{equation}
where $c_{sn}$ and $k_\mathrm{sn}$ are, respectively, the fatigue strength and  slope of the S-N curve.

Comparing the Basquin's damage model  with the model described by (\ref{eq:semikron}) for semiconductor switches, the power law relationship  approximately holds for the temperature swing $\Delta T_j$ but not for the other two stress factors, i.e., the average temperature and the cycle duration. In the following we will work under the realistic assumption that  the temperature variation is the dominant stress factor and neglect the effect of the other two factors variation  on the number of available life-cycles and hence on the damage. 

\subsection{Damage estimation in  frequency-domain}
\label{sec:freqD}
The damage formula defined in  (\ref{eq:total_rainflow}), requires the knowledge  of the effective number of cycles $n_i(S_i)$ at stress level $S_i$. This is obtained by counting cycles of different stress level in the stress history. In practice  cycle counting algorithms, such as the Rainflow\cite{rainflow}, are employed.  As an alternative  to those  traditional time-domain nonlinear approaches, spectral methods in  frequency domain are employed  to estimate $n_i(S_i)$  and thus fatigue damage 
 without the drawbacks of counting algorithms such as computational complexity.
 
In this work, we will employ  the frequency domain based ``single moment'' method described in \cite{single_moment} to  model the damage.
This method considers the stress load signal as  a uniaxial stochastic process, and  estimates the
fatigue induced  damage as a sum of damage
contributions of narrow-band processes, defined by decomposing the
original power spectrum in to infinitesimal narrow-band PSDs with  bandwidth of $\Delta \omega$ (see \ref{fig:fatigue}b).
In \cite{nasa_comparison}, authors show that, for small values of the S-N exponent ($k_{sn}\sim3$),   the single moment method provide a good approximation of the damage obtained via the Rainflow based time-domain  approach for most cases. 

Using the single moment spectral approach, the total damage is then computed  as 
  \begin{equation}
  \label{eq:SM}
      D_\mathrm{tot}=\left(\sum^{N}_{i=1}D^{\frac{2}{k_{\mathrm{sn}}}}_i\right)^\frac{k_{\mathrm{sn}}}{2}
  \end{equation}
where the damage $D_i$ is interpreted as contribution of the infinitesimal power spectrum at around frequency $\omega_i$:

\begin{equation}
    D_i=\frac{\omega_i}{2\pi c_{sn}}\left(\sqrt{2S(\omega_i)\Delta \omega}\right)^{k_\mathrm{sn}} \Gamma\left(1+\frac{k_\mathrm{sn}}{2}\right),
\end{equation}
where $S(\omega)$ is the PSD of the stress signal, which in the case of semiconductor damage,  would be the temperature signal, and $\Gamma$ is the gamma function.

\begin{figure}[h] 
\centering
\includegraphics[width=0.8\linewidth]{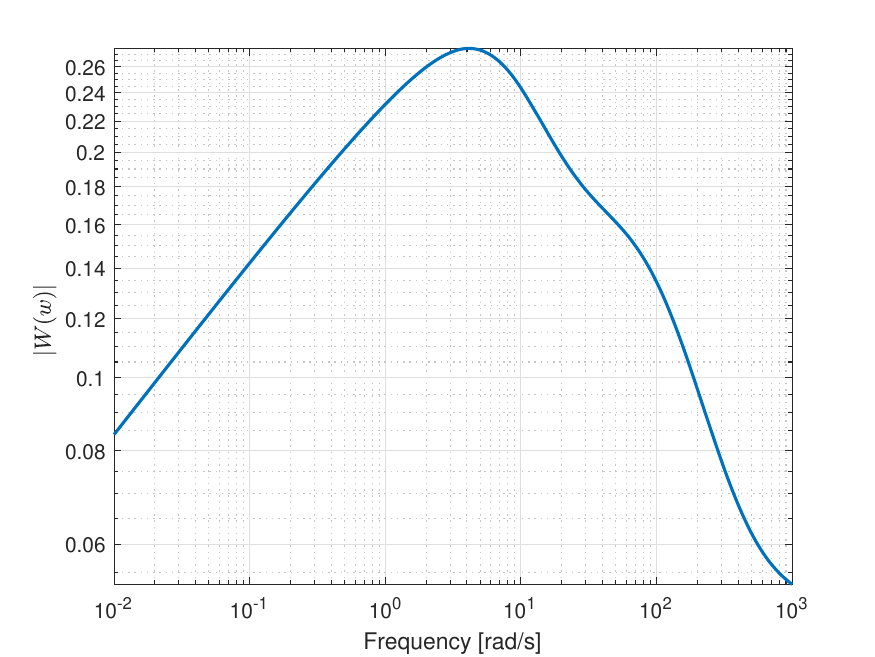}
\caption{ The $W(\omega)$ evaluated at different frequencies, reflecting the contribution of damage at each frequency. \label{fig:W}}
\end{figure}

The total damage can be then reformulated as follows (see details in \cite{Amin_short_note}):
  \begin{equation}
  \label{eq:Dtot}
      D_\mathrm{tot}\propto\left(\sum^{N}_{i=1}   \left|W(\omega_i)\right|^2 \left|G_{P,i_{ref}}(\omega_i)\right|^2 S_{i_{ref}}(\omega_i)\Delta \omega  \right)^\frac{\alpha}{2},
  \end{equation}
  where  $S_{i_{ref}}$ denotes the PSD of the input reference current, $G_{P,i_{ref}}$ is the closed-loop transfer function from the reference current to the generated power at the semiconductor, and $W$ is a frequency-dependent weight function as plotted in Fig. \ref{fig:W}.
As we can see, at low frequencies the damage accumulation is insignificant because of the reduced  number of cycles at these frequencies. At higher frequencies, the number of cycles is limited  due to the bandwidth of the thermodynamics.

\section{Reliability Control Design}
\label{sec:control}
As illustrated in Fig. \ref{fig:system} the current controller operates in the rotating dq-coordinate frame with the objective of tracking the desired current signals. As we observe from the   system dynamics described in (\ref{eq:dynamics}), there is a coupling between d and q components of the current. Furthermore, the frequency $\omega$ can be time-varying and therefore incorporating the frequency variable into the control design framework can become cumbersome. Therefore, we decouple the $i_d$ and $i_q$ dynamics by introducing an auxiliary control variable $u$ which is defined as:
\begin{eqnarray}
\label{eq:aux}
    u_d:&=&v_{td}-v_d+\omega L_f i_q, \nonumber \\
    u_q:&=&v_{tq}-v_q-\omega L_f i_d.
\end{eqnarray}

With this definition, the system dynamics become decoupled and independent of $\omega$, resulting in a linear time-invariant (LTI)  system. Once the controller outputs, i.e. $u_d$ and $u_q$,  are computed, the input to the modulation system can be readily determined by solving (\ref{eq:aux}) for $v_{td}$ and $v_{tq}$ (see Fig. \ref{fig:system} for the signal flow). A potential problem is that the computed  modulation inputs can become infeasible (over-modulation) and, as a result, the closed-loop system becomes nonlinear.

In the following, we attempt to design a controller for the q-component of the current, the derivation for the d-component controller is analog. 
As discussed in Section \ref{sec:freqD}, the semiconductor fatigue due to thermal stress is influenced by the frequency contents mostly in the range of 1 to 100 rad/s. Therefore, the objective of the reliability control design is to attenuate the frequency contents within this band gap, without compromising the current tracking requirement. In other words, we aim for a controller capable of tracking the reference as fast as possible, without causing significant damage. 

\begin{figure}[h]
  \centering
  \includegraphics[width=0.8\linewidth]{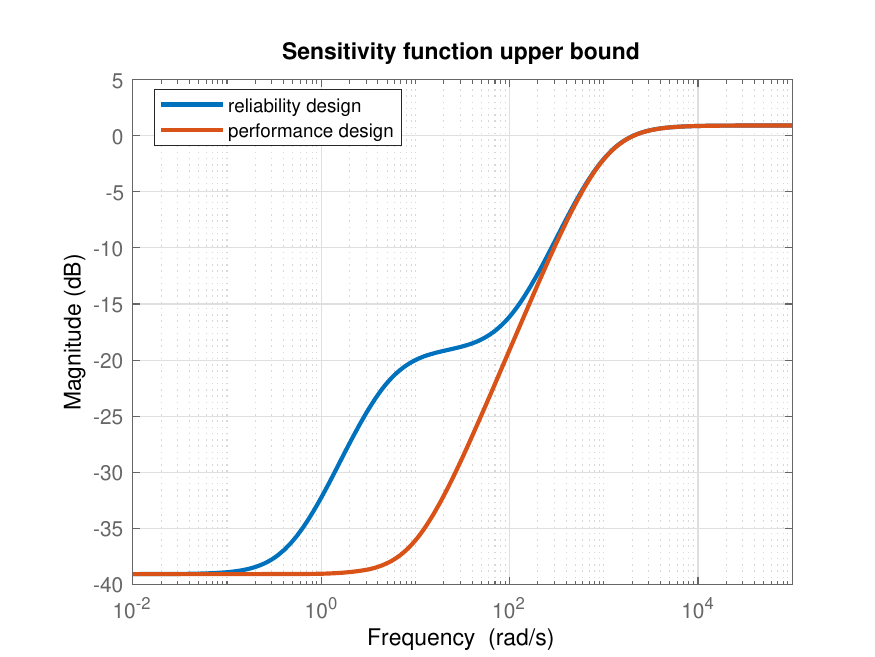}
  \caption{The sensitivity function upper bound for the two control modes: the performance (red) and reliability (blue) designs. \label{fig:weights}}
\end{figure}

This problem can be tackled by defining an upper bound on the sensitivity function of the closed-loop system and then design a linear controller such that the resulting sensitivity function is below this threshold at each frequency. A typical frequency-dependent upper-bound for the sensitivity function is plotted in red in Fig. \ref{fig:weights}, which consists of a pole and a zero. From the final value theorem (FVT), the value of the bound at the DC frequency denotes the upper bound on the steady-state tracking error. The frequency of the zero determines the bandwidth requirement of the closed-loop tracking transfer function. Due to the water-bed effect in linear control systems, the sensitivity function cannot be pushed down at all frequencies. That is, for example, to achieve lower steady-state error, the transient behavior should be compromised, i.e. pushing up the sensitivity function at higher frequencies. To reduce the damage to the semiconductor switches, we allow the sensitivity upper bound to slightly increase by adding one pole and one zero around the band gap of 1 to 100 rad/sec, the range where we analysed to have the most damage. This upper bound is plotted in blue in Fig. \ref{fig:weights}, and the corresponding design will be addressed as the ``reliability design'' in the sequel, compared to the previous upper bound which will be called the ``performance design'', as it has tighter requirement in reference tracking.

As stated earlier, the magnitude of the closed-loop  sensitivity function $S(K, j\omega)$  is required to be upper bounded by$W(j\omega)$ at each frequency. Mathematically speaking, this can be expressed as
\begin{equation}\label{eq:upperbound}
    \left|S(K, j\omega)\right|\leq \left| W(j\omega)\right|, ~~~~\textrm{for all $\omega$} 
\end{equation}
where  $K$ denotes the controller to be designed.

The inequality above is equivalent to the $\mathcal{H}_{\infty}$ norm, which, in this case, is the peak magnitude of the transfer function, given that this is a single-input single-output (SISO) system. In other words,
\begin{equation}
        \|  W(j\omega)S(K, j\omega)\|_{\infty} \leq 1. 
\end{equation}

Finding the controller $K(j\omega)$ such that it fulfils the above constraint is carried out by the $\mathcal{H}_{\infty}$ control synthesis method:

\begin{align}
\underset{\gamma, ~ K(s)}{\text{minimize}} & ~~~\gamma\nonumber \\
\textrm{subject to } ~ &\|  W(j\omega)S(K, j\omega)\|_{\infty} \leq \gamma.
\end{align}

Note that the resulting controller is a MIMO system as it takes the current measurement, i.e. $i_q$, and the reference value $i_{q,ref}$, and computes the $u_q$ as the output. 

\section{Simulation Results}
\label{sec:simulation}
\subsection{Current tracking performance}
\begin{figure}[h]
  \centering
  \includegraphics[width=0.8\linewidth]{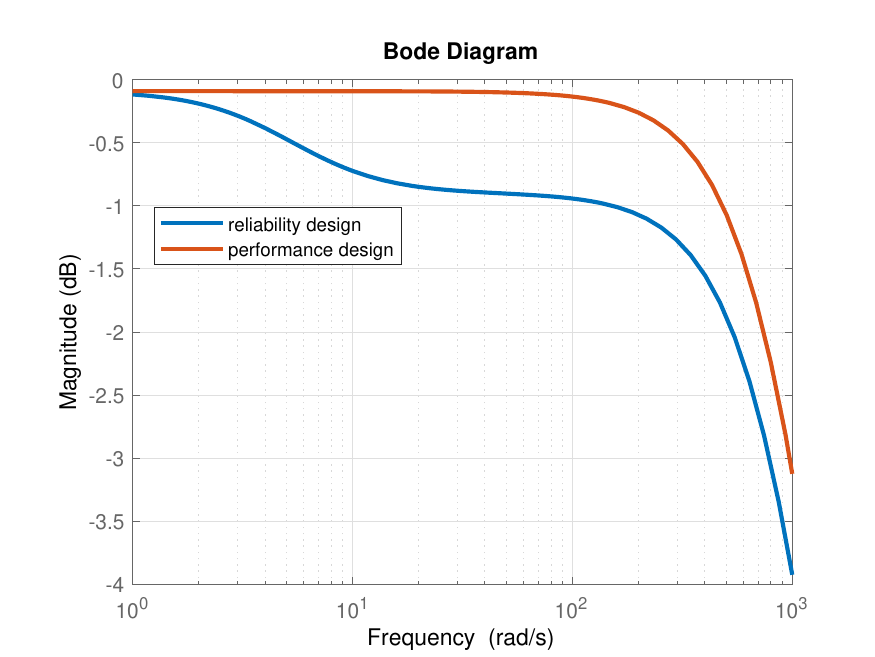} 
  \caption{The Bode diagram of the closed-loop transfer function (from current reference to the output current) for both design upper bounds.\label{fig:bodePc}}
\end{figure}

\begin{figure}[h]
  \centering
  \includegraphics[width=0.8\linewidth]{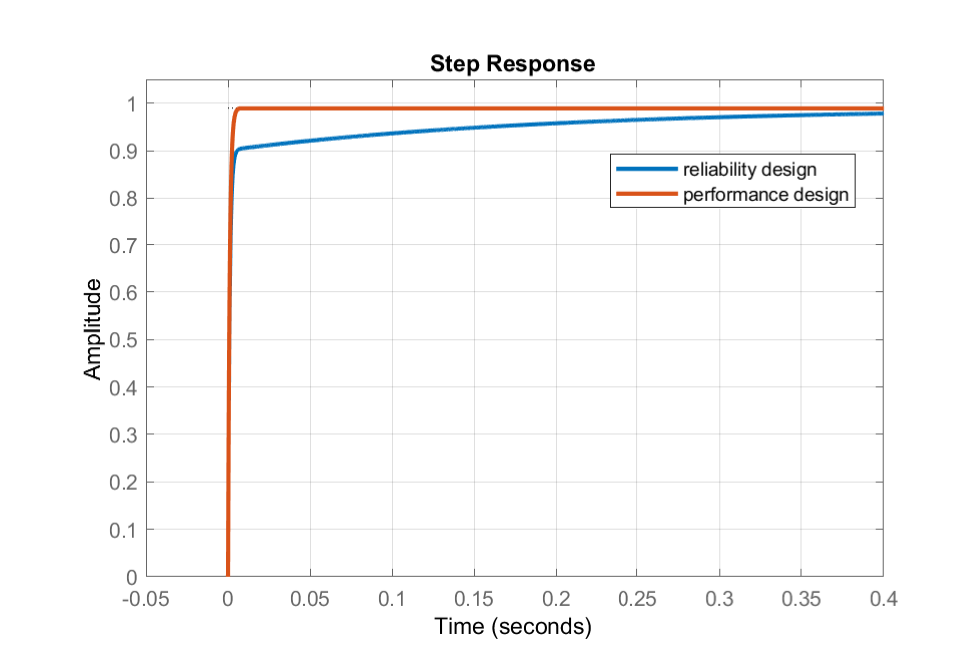} 
  \caption{The closed-loop step response of the two design upper bounds.\label{fig:step}}
\end{figure}

The resulting closed-loop system transfer functions for both upper bounds are plotted in Fig. \ref{fig:bodePc}. As we can see, with the reliability design, the closed-loop transfer function attenuates slightly the input frequency contents around the region where the damage spectrum has highest gain. Both design have approximately the same closed-loop bandwidth. In terms of of the time-domain response, both designs can reach the reference point at approximately the same speed. This can be observed in the step response of the closed-loop system illustrated in Fig. \ref{fig:step}. As we can see, both controllers achieve the same rise-time, reaching 90\% of the final value at approximately the same time.  However, the settling-time of the reliability design is much longer, and this is because of bending the sensitivity upper bound at around 0.1 rad/sec to gain reliability and semiconductor lifetime.

\subsection{Lifetime assessment}
\begin{figure}[h]
  \centering
  \includegraphics[width=0.8\linewidth]{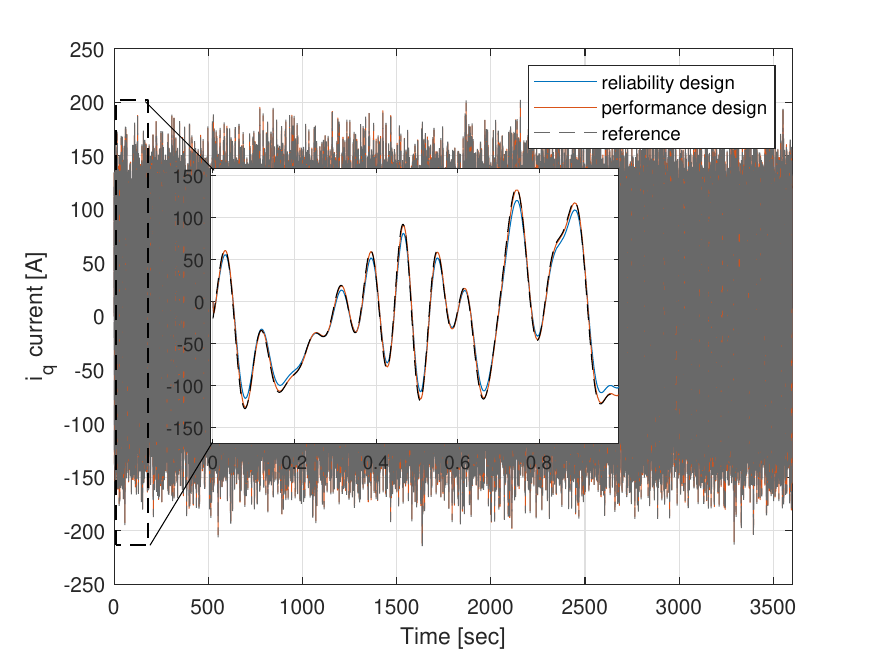}
  \caption{The inverter output current profiles: the reference (dashed-line), the current profile using the performance design (red), and the reliability design (blue).\label{fig:Iq}}
\end{figure}

\begin{figure}[h]
  \centering
  \includegraphics[width=0.8\linewidth]{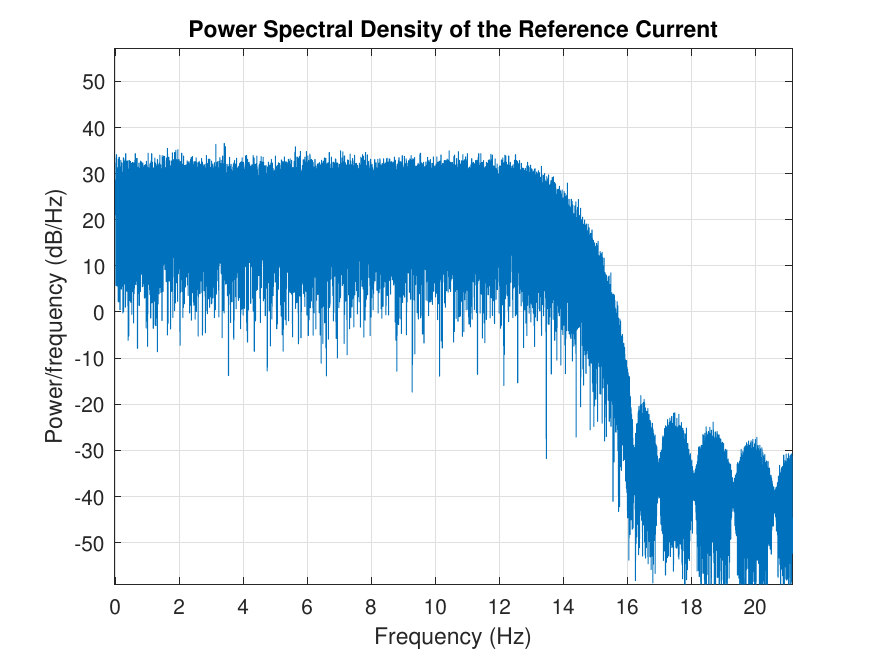}
  \caption{The power spectral density of the reference current profile.\label{fig:Iqpsd}}
\end{figure}

\begin{figure}[h]
  \centering
  \includegraphics[width=0.8\linewidth]{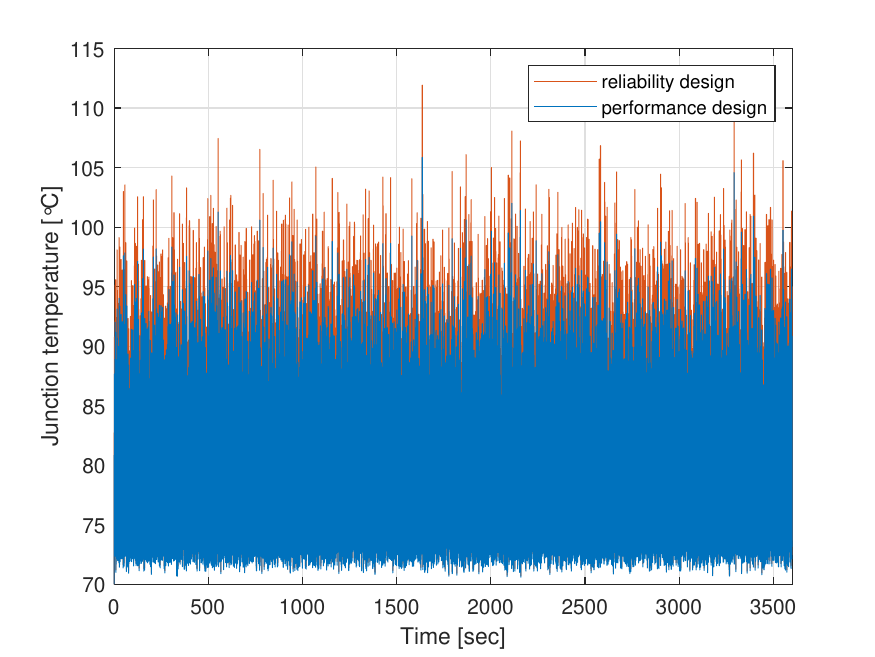}
  \caption{The junction temperature of the SiC MOSFET: with the reliability design (blue) and the performance design (red).\label{fig:Tj}}
\end{figure}

To evaluate the impact of the controller  dynamics on the accumulated damage, we simulate the time domain response of the closed-loop system. The reference current profile is chosen to be a stochastic process of length 1 hour, as shown in Fig.  \ref{fig:Iq}, and the sampling time is set to 1 milliseconds. The power spectral density (PSD) of the reference current profile is plotted in Fig. \ref{fig:Iqpsd}. The bandwidth of the reference signal is chosen to be as wide as the damage band gap, i.e. in the range of 100 rad/sec. The output current profiles of the two control designs are plotted in Fig. \ref{fig:Iq}. As we can see, using the reliability control design leads to slightly higher tracking error, however, when it comes to comparing the thermal stress, as depicted in Fig. \ref{fig:Tj}, it leads to lower temperature swing and  median temperature, which are influential in the device lifetime.

\subsubsection{Monte Carlo simulation}
\begin{figure}[h]
  \centering
  \includegraphics[width=0.9\linewidth]{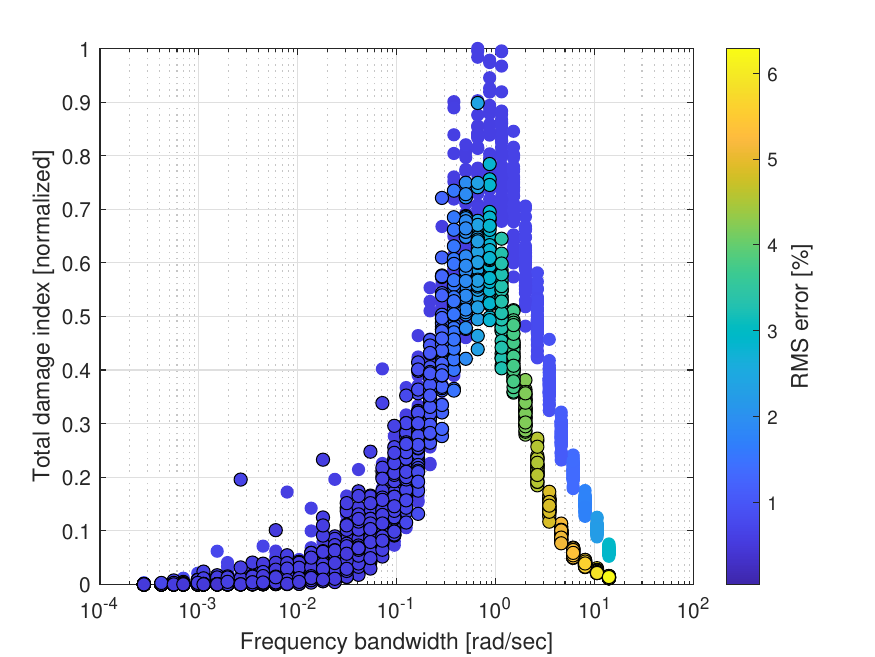}
  \caption{The total damage index, computed by the Rainflow counting method, versus different PSD bandwidths of the reference current. The circles with black outline represent the reliability design.  The damage is normalized with respect to the maximum damage accumulated by the performance design.  \label{fig:MontCarlo}}
\end{figure}

To assess the performance of both control designs, in term of improved converter lifetime,  a series of simulations were conducted, each with a random stochastic input current of length 3600 seconds and standard deviation of 40A. In addition, the bandwidth of the reference PSD is also varied to study the frequency response. Figure \ref{fig:MontCarlo} illustrates the damage index versus different input PSD bandwidth. The damage is normalized with respect to the maximum accumulated damage of the controller with performance design. The result of the reliability design is depicted with circles with black outline. As we can see, with this control design, the total damage is relatively lower specially in the frequency gap of 1 to 10rad/s. However, this comes at the cost of slightly higher tracking error as depicted by the colors filling the circles  denoting the normalized RMS error. 

\section{Conclusions}
\label{sec:conc}
This paper employs the concept of reliability-optimized control, whereby the operation of a power converter is optimized to minimize the damage and maximizing the converter's operational lifetime.
Specifically, we proposed a control policy that integrates the tracking and reliability requirements in an optimization framework, thereby enabling the efficient and reliable operation of power converters under varying  conditions.
The effectiveness of the proposed control policy was illustrated through simulations, which demonstrated the benefits of the approach in limiting damage to power semiconductor devices induced by thermal cycling while fulfilling the current tracking requirements.
The proposed approach offers a novel perspective and insights into the sustainable operation of power electronics converters, paving the way for a comprehensive suite of studies towards a more environmentally aware and sustainable future in power conversion.

\bibliographystyle{ACM-Reference-Format}
\bibliography{reference}


\begin{thebibliography}{47}


\ifx \showCODEN    \undefined \def \showCODEN     #1{\unskip}     \fi
\ifx \showDOI      \undefined \def \showDOI       #1{#1}\fi
\ifx \showISBNx    \undefined \def \showISBNx     #1{\unskip}     \fi
\ifx \showISBNxiii \undefined \def \showISBNxiii  #1{\unskip}     \fi
\ifx \showISSN     \undefined \def \showISSN      #1{\unskip}     \fi
\ifx \showLCCN     \undefined \def \showLCCN      #1{\unskip}     \fi
\ifx \shownote     \undefined \def \shownote      #1{#1}          \fi
\ifx \showarticletitle \undefined \def \showarticletitle #1{#1}   \fi
\ifx \showURL      \undefined \def \showURL       {\relax}        \fi
\providecommand\bibfield[2]{#2}
\providecommand\bibinfo[2]{#2}
\providecommand\natexlab[1]{#1}
\providecommand\showeprint[2][]{arXiv:#2}

\bibitem[Aly et~al\mbox{.}(2017)]%
        {optimal_svm}
\bibfield{author}{\bibinfo{person}{Mokhtar Aly}, \bibinfo{person}{Emad~M. Ahmed}, {and} \bibinfo{person}{Masahito Shoyama}.} \bibinfo{year}{2017}\natexlab{}.
\newblock \showarticletitle{Developing new lifetime prolongation SVM algorithm for multilevel inverters with thermally aged power devices}.
\newblock \bibinfo{journal}{\emph{IET Power Electronics}} \bibinfo{volume}{10}, \bibinfo{number}{15} (\bibinfo{year}{2017}), \bibinfo{pages}{2248--2256}.
\newblock


\bibitem[ASTM et~al\mbox{.}(2011)]%
        {rainflow}
\bibfield{author}{\bibinfo{person}{E ASTM} {et~al\mbox{.}}} \bibinfo{year}{2011}\natexlab{}.
\newblock \showarticletitle{Standard practices for cycle counting in fatigue analysis}.
\newblock \bibinfo{journal}{\emph{Standard Practices for Cycle Counting in Fatigue Analysis}} (\bibinfo{year}{2011}).
\newblock


\bibitem[Bakran and Knaak(2003)]%
        {bakran2003aktive}
\bibfield{author}{\bibinfo{person}{M Bakran} {and} \bibinfo{person}{H Knaak}.} \bibinfo{year}{2003}\natexlab{}.
\newblock \showarticletitle{Aktive stabilisierung des temperaturniveaus in halbleiterbauelementen}.
\newblock \bibinfo{journal}{\emph{DE2}} \bibinfo{volume}{2}, \bibinfo{number}{103} (\bibinfo{year}{2003}), \bibinfo{pages}{790}.
\newblock


\bibitem[Barresi et~al\mbox{.}(2024)]%
        {thermalControl5}
\bibfield{author}{\bibinfo{person}{Marzio Barresi}, \bibinfo{person}{Luigi Piegari}, {and} \bibinfo{person}{Riccardo Scalabrin}.} \bibinfo{year}{2024}\natexlab{}.
\newblock \showarticletitle{Thermal Adaptive Space Vector Modulation for Current Source Converters}. In \bibinfo{booktitle}{\emph{2024 International Symposium on Power Electronics, Electrical Drives, Automation and Motion (SPEEDAM)}}. \bibinfo{pages}{895--902}.
\newblock
\urldef\tempurl%
\url{https://doi.org/10.1109/SPEEDAM61530.2024.10609191}
\showDOI{\tempurl}


\bibitem[Benasciutti et~al\mbox{.}(2013)]%
        {single_moment}
\bibfield{author}{\bibinfo{person}{D. Benasciutti}, \bibinfo{person}{A. Cristofori}, {and} \bibinfo{person}{R. Tovo}.} \bibinfo{year}{2013}\natexlab{}.
\newblock \showarticletitle{Analogies between spectral methods and multiaxial criteria in fatigue damage evaluation}.
\newblock \bibinfo{journal}{\emph{Probabilistic Engineering Mechanics}}  \bibinfo{volume}{31} (\bibinfo{year}{2013}), \bibinfo{pages}{39--45}.
\newblock
\showISSN{0266-8920}


\bibitem[Blasko et~al\mbox{.}(1999)]%
        {801687}
\bibfield{author}{\bibinfo{person}{V. Blasko}, \bibinfo{person}{R. Lukaszewski}, {and} \bibinfo{person}{R. Sladky}.} \bibinfo{year}{1999}\natexlab{}.
\newblock \showarticletitle{On line thermal model and thermal management strategy of a three phase voltage source inverter}. In \bibinfo{booktitle}{\emph{Conference Record of the 1999 IEEE Industry Applications Conference. Thirty-Forth IAS Annual Meeting (Cat. No.99CH36370)}}, Vol.~\bibinfo{volume}{2}. \bibinfo{pages}{1423--1431 vol.2}.
\newblock
\urldef\tempurl%
\url{https://doi.org/10.1109/IAS.1999.801687}
\showDOI{\tempurl}


\bibitem[Chen et~al\mbox{.}(2024)]%
        {reliability_switching}
\bibfield{author}{\bibinfo{person}{Yiyang Chen}, \bibinfo{person}{Yimin Zhang}, \bibinfo{person}{Haoyu Chen}, \bibinfo{person}{Zhen Li}, {and} \bibinfo{person}{Zhenbin Zhang}.} \bibinfo{year}{2024}\natexlab{}.
\newblock \showarticletitle{Anti-Fatigue-Damage-Oriented Through-Life Optimization and Control of High-Power IGCT Converters in Wind Energy Systems}.
\newblock \bibinfo{journal}{\emph{Symmetry}} \bibinfo{volume}{16}, \bibinfo{number}{8} (\bibinfo{year}{2024}).
\newblock
\showISSN{2073-8994}
\urldef\tempurl%
\url{https://doi.org/10.3390/sym16081047}
\showDOI{\tempurl}


\bibitem[Ciappa and Fichtner(2000)]%
        {bondwire2}
\bibfield{author}{\bibinfo{person}{M. Ciappa} {and} \bibinfo{person}{W. Fichtner}.} \bibinfo{year}{2000}\natexlab{}.
\newblock \showarticletitle{Lifetime prediction of IGBT modules for traction applications}. In \bibinfo{booktitle}{\emph{2000 IEEE International Reliability Physics Symposium Proceedings}}. \bibinfo{pages}{210--216}.
\newblock


\bibitem[Czechowski and Lenk(1978)]%
        {Miner1}
\bibfield{author}{\bibinfo{person}{Antoni Czechowski} {and} \bibinfo{person}{Arno Lenk}.} \bibinfo{year}{1978}\natexlab{}.
\newblock \showarticletitle{Miner's Rule in Mechanical Tests of Electronic Parts}.
\newblock \bibinfo{journal}{\emph{IEEE Transactions on Reliability}} \bibinfo{volume}{R-27}, \bibinfo{number}{3} (\bibinfo{year}{1978}), \bibinfo{pages}{183--190}.
\newblock
\urldef\tempurl%
\url{https://doi.org/10.1109/TR.1978.5220318}
\showDOI{\tempurl}


\bibitem[di~Benedetto et~al\mbox{.}(2020)]%
        {fault_tol2}
\bibfield{author}{\bibinfo{person}{Marco di Benedetto}, \bibinfo{person}{Alessandro Lidozzi}, \bibinfo{person}{Luca Solero}, \bibinfo{person}{Fabio Crescimbini}, {and} \bibinfo{person}{Petar~J. Grbović}.} \bibinfo{year}{2020}\natexlab{}.
\newblock \showarticletitle{Reliability and Real-Time Failure Protection of the Three-Phase Five-Level E-Type Converter}.
\newblock \bibinfo{journal}{\emph{IEEE Transactions on Industry Applications}} \bibinfo{volume}{56}, \bibinfo{number}{6} (\bibinfo{year}{2020}), \bibinfo{pages}{6630--6641}.
\newblock
\urldef\tempurl%
\url{https://doi.org/10.1109/TIA.2020.3019358}
\showDOI{\tempurl}


\bibitem[Ghazanfari and Mohamed(2016)]%
        {redundant}
\bibfield{author}{\bibinfo{person}{Amin Ghazanfari} {and} \bibinfo{person}{Yasser Abdel-Rady~I. Mohamed}.} \bibinfo{year}{2016}\natexlab{}.
\newblock \showarticletitle{A Resilient Framework for Fault-Tolerant Operation of Modular Multilevel Converters}.
\newblock \bibinfo{journal}{\emph{IEEE Transactions on Industrial Electronics}} \bibinfo{volume}{63}, \bibinfo{number}{5} (\bibinfo{year}{2016}), \bibinfo{pages}{2669--2678}.
\newblock
\urldef\tempurl%
\url{https://doi.org/10.1109/TIE.2016.2516968}
\showDOI{\tempurl}


\bibitem[Gutsch and Leker(2022)]%
        {GUTSCH2022105030}
\bibfield{author}{\bibinfo{person}{Moritz Gutsch} {and} \bibinfo{person}{Jens Leker}.} \bibinfo{year}{2022}\natexlab{}.
\newblock \showarticletitle{Global warming potential of lithium-ion battery energy storage systems: A review}.
\newblock \bibinfo{journal}{\emph{Journal of Energy Storage}}  \bibinfo{volume}{52} (\bibinfo{year}{2022}), \bibinfo{pages}{105030}.
\newblock
\showISSN{2352-152X}
\urldef\tempurl%
\url{https://doi.org/10.1016/j.est.2022.105030}
\showDOI{\tempurl}


\bibitem[Hernes et~al\mbox{.}(2021)]%
        {rel}
\bibfield{author}{\bibinfo{person}{Magnar Hernes}, \bibinfo{person}{Salvatore D'Arco}, \bibinfo{person}{Antonios Antonopoulos}, {and} \bibinfo{person}{Dimosthenis Peftitsis}.} \bibinfo{year}{2021}\natexlab{}.
\newblock \showarticletitle{Failure analysis and lifetime assessment of IGBT power modules at low temperature stress cycles}.
\newblock \bibinfo{journal}{\emph{IET Power Electronics}} \bibinfo{volume}{14}, \bibinfo{number}{7} (\bibinfo{year}{2021}), \bibinfo{pages}{1271--1283}.
\newblock


\bibitem[{Kumar} et~al\mbox{.}(2021)]%
        {fault_tol1}
\bibfield{author}{\bibinfo{person}{Dhananjay {Kumar}}, \bibinfo{person}{Rajesh~Kumar {Nema}}, {and} \bibinfo{person}{Sushma {Gupta}}.} \bibinfo{year}{2021}\natexlab{}.
\newblock \showarticletitle{{Investigation of fault-tolerant capabilities of some recent multilevel inverter topologies}}.
\newblock \bibinfo{journal}{\emph{International Journal of Electronics}} \bibinfo{volume}{108}, \bibinfo{number}{11} (\bibinfo{date}{Nov.} \bibinfo{year}{2021}), \bibinfo{pages}{1957--1976}.
\newblock
\urldef\tempurl%
\url{https://doi.org/10.1080/00207217.2020.1870752}
\showDOI{\tempurl}


\bibitem[Larsen and Irvine(2015)]%
        {nasa_comparison}
\bibfield{author}{\bibinfo{person}{Curtis~E. Larsen} {and} \bibinfo{person}{Tom Irvine}.} \bibinfo{year}{2015}\natexlab{}.
\newblock \showarticletitle{A Review of Spectral Methods for Variable Amplitude Fatigue Prediction and New Results}.
\newblock \bibinfo{journal}{\emph{Procedia Engineering}}  \bibinfo{volume}{101} (\bibinfo{year}{2015}), \bibinfo{pages}{243--250}.
\newblock
\showISSN{1877-7058}


\bibitem[Lewis and Grainger(2020)]%
        {thermalControl4}
\bibfield{author}{\bibinfo{person}{Patrick~T. Lewis} {and} \bibinfo{person}{Brandon~M. Grainger}.} \bibinfo{year}{2020}\natexlab{}.
\newblock \showarticletitle{Preserving Converter Lifetime by Active Thermal Boundary Control}. In \bibinfo{booktitle}{\emph{2020 IEEE Energy Conversion Congress and Exposition (ECCE)}}. \bibinfo{pages}{5310--5317}.
\newblock
\urldef\tempurl%
\url{https://doi.org/10.1109/ECCE44975.2020.9236215}
\showDOI{\tempurl}


\bibitem[Li et~al\mbox{.}(2014)]%
        {cooling1}
\bibfield{author}{\bibinfo{person}{Cong Li}, \bibinfo{person}{Da Jiao}, \bibinfo{person}{Jizhou Jia}, \bibinfo{person}{Feng Guo}, {and} \bibinfo{person}{Jin Wang}.} \bibinfo{year}{2014}\natexlab{}.
\newblock \showarticletitle{Thermoelectric Cooling for Power Electronics Circuits: Modeling and Active Temperature Control}.
\newblock \bibinfo{journal}{\emph{IEEE Transactions on Industry Applications}} \bibinfo{volume}{50}, \bibinfo{number}{6} (\bibinfo{year}{2014}), \bibinfo{pages}{3995--4005}.
\newblock
\urldef\tempurl%
\url{https://doi.org/10.1109/TIA.2014.2319576}
\showDOI{\tempurl}


\bibitem[Lorenz(2002)]%
        {997541}
\bibfield{author}{\bibinfo{person}{R.D. Lorenz}.} \bibinfo{year}{2002}\natexlab{}.
\newblock \showarticletitle{Power conversion challenges with a multidisciplinary focus}. In \bibinfo{booktitle}{\emph{Proceedings of the Power Conversion Conference-Osaka 2002 (Cat. No.02TH8579)}}, Vol.~\bibinfo{volume}{2}. \bibinfo{pages}{347--352 vol.2}.
\newblock
\urldef\tempurl%
\url{https://doi.org/10.1109/PCC.2002.997541}
\showDOI{\tempurl}


\bibitem[Mohan et~al\mbox{.}(2003)]%
        {mohan}
\bibfield{author}{\bibinfo{person}{Ned Mohan}, \bibinfo{person}{Tore~M. Undeland}, {and} \bibinfo{person}{William~P. Robbins}.} \bibinfo{year}{2003}\natexlab{}.
\newblock \bibinfo{booktitle}{\emph{Power Electronics. Converters, Applications and Design} (\bibinfo{edition}{third} ed.)}.
\newblock \bibinfo{publisher}{John Wiley and Sons, Inc}.
\newblock


\bibitem[Murdock and Lorenz(2002)]%
        {Murdock}
\bibfield{author}{\bibinfo{person}{D.A. Murdock} {and} \bibinfo{person}{R.D. Lorenz}.} \bibinfo{year}{April 2002}\natexlab{}.
\newblock \showarticletitle{Failure Minimizing, active Control of IPEMs}. In \bibinfo{booktitle}{\emph{Proc of 2002 CPES Seminar}}, Vol.~\bibinfo{volume}{2}.
\newblock
\urldef\tempurl%
\url{https://doi.org/10.1109/IAS.1999.801687}
\showDOI{\tempurl}


\bibitem[Murdock et~al\mbox{.}(2006a)]%
        {cooling2}
\bibfield{author}{\bibinfo{person}{D.A. Murdock}, \bibinfo{person}{J.E.R. Torres}, \bibinfo{person}{J.J. Connors}, {and} \bibinfo{person}{R.D. Lorenz}.} \bibinfo{year}{2006}\natexlab{a}.
\newblock \showarticletitle{Active thermal control of power electronic modules}.
\newblock \bibinfo{journal}{\emph{IEEE Transactions on Industry Applications}} \bibinfo{volume}{42}, \bibinfo{number}{2} (\bibinfo{year}{2006}), \bibinfo{pages}{552--558}.
\newblock
\urldef\tempurl%
\url{https://doi.org/10.1109/TIA.2005.863905}
\showDOI{\tempurl}


\bibitem[Murdock et~al\mbox{.}(2006b)]%
        {Murdock2}
\bibfield{author}{\bibinfo{person}{D.A. Murdock}, \bibinfo{person}{J.E.R. Torres}, \bibinfo{person}{J.J. Connors}, {and} \bibinfo{person}{R.D. Lorenz}.} \bibinfo{year}{2006}\natexlab{b}.
\newblock \showarticletitle{Active thermal control of power electronic modules}.
\newblock \bibinfo{journal}{\emph{IEEE Transactions on Industry Applications}} \bibinfo{volume}{42}, \bibinfo{number}{2} (\bibinfo{year}{2006}), \bibinfo{pages}{552--558}.
\newblock
\urldef\tempurl%
\url{https://doi.org/10.1109/TIA.2005.863905}
\showDOI{\tempurl}


\bibitem[Naveen and Ms.Hema~Rani(2021)]%
        {thermalControl3}
\bibfield{author}{\bibinfo{person}{Pogula Naveen} {and} \bibinfo{person}{P Ms.Hema~Rani}.} \bibinfo{year}{2021}\natexlab{}.
\newblock \showarticletitle{Reliability Improvement of IGBT Employed in Boost Converter using Active Thermal Control}. In \bibinfo{booktitle}{\emph{2021 International Conference on Intelligent Technologies (CONIT)}}. \bibinfo{pages}{1--6}.
\newblock
\urldef\tempurl%
\url{https://doi.org/10.1109/CONIT51480.2021.9498456}
\showDOI{\tempurl}


\bibitem[Novak and Blaabjerg(2020)]%
        {thermalControl2}
\bibfield{author}{\bibinfo{person}{Mateja Novak} {and} \bibinfo{person}{Frede Blaabjerg}.} \bibinfo{year}{2020}\natexlab{}.
\newblock \showarticletitle{Model Predictive Active Thermal Control Strategy for Lifetime Extension of a 3L-NPC Converter for UPS Applications}. In \bibinfo{booktitle}{\emph{2020 IEEE 21st Workshop on Control and Modeling for Power Electronics (COMPEL)}}. \bibinfo{pages}{1--7}.
\newblock
\urldef\tempurl%
\url{https://doi.org/10.1109/COMPEL49091.2020.9265807}
\showDOI{\tempurl}


\bibitem[of~Defense(1995)]%
        {MIL-HDBK-217F}
\bibfield{editor}{\bibinfo{person}{US~Department of Defense}} (Ed.). \bibinfo{year}{1995}\natexlab{}.
\newblock \bibinfo{booktitle}{\emph{The Military Handbook for Reliability Prediction of Electronic Equipment MIL-HDBK-217F}}. \bibinfo{publisher}{Defense Printing Service}.
\newblock


\bibitem[Phan et~al\mbox{.}(2014)]%
        {svm2}
\bibfield{author}{\bibinfo{person}{The-Minh Phan}, \bibinfo{person}{Nikolaos Oikonomou}, \bibinfo{person}{Gernot~J. Riedel}, {and} \bibinfo{person}{Mario Pacas}.} \bibinfo{year}{2014}\natexlab{}.
\newblock \showarticletitle{PWM for active thermal protection in three level neutral point clamped inverters}. In \bibinfo{booktitle}{\emph{2014 IEEE Energy Conversion Congress and Exposition (ECCE)}}. \bibinfo{pages}{3710--3716}.
\newblock
\urldef\tempurl%
\url{https://doi.org/10.1109/ECCE.2014.6953905}
\showDOI{\tempurl}


\bibitem[Phan et~al\mbox{.}(2015)]%
        {svm3}
\bibfield{author}{\bibinfo{person}{The-Minh Phan}, \bibinfo{person}{Gernot~J. Riedel}, \bibinfo{person}{Nikolaos Oikonomou}, {and} \bibinfo{person}{Mario Pacas}.} \bibinfo{year}{2015}\natexlab{}.
\newblock \showarticletitle{Active thermal protection and lifetime extension in 3L-NPC-inverter in the low modulation range}. In \bibinfo{booktitle}{\emph{2015 IEEE Applied Power Electronics Conference and Exposition (APEC)}}. \bibinfo{pages}{2269--2276}.
\newblock
\urldef\tempurl%
\url{https://doi.org/10.1109/APEC.2015.7104665}
\showDOI{\tempurl}


\bibitem[Popović-Gerber et~al\mbox{.}(2012)]%
        {PEC1}
\bibfield{author}{\bibinfo{person}{Jelena Popović-Gerber}, \bibinfo{person}{Jesus~Angel Oliver}, \bibinfo{person}{Nicolás Cordero}, \bibinfo{person}{Thomas Harder}, \bibinfo{person}{José~A. Cobos}, \bibinfo{person}{Michael Hayes}, \bibinfo{person}{Seán~Cian O’Mathuna}, {and} \bibinfo{person}{Erich Prem}.} \bibinfo{year}{2012}\natexlab{}.
\newblock \showarticletitle{Power Electronics Enabling Efficient Energy Usage: Energy Savings Potential and Technological Challenges}.
\newblock \bibinfo{journal}{\emph{IEEE Transactions on Power Electronics}} \bibinfo{volume}{27}, \bibinfo{number}{5} (\bibinfo{year}{2012}), \bibinfo{pages}{2338--2353}.
\newblock
\urldef\tempurl%
\url{https://doi.org/10.1109/TPEL.2011.2171195}
\showDOI{\tempurl}


\bibitem[Qin et~al\mbox{.}(2013)]%
        {SiC1}
\bibfield{author}{\bibinfo{person}{Haihong Qin}, \bibinfo{person}{Bin Zhao}, \bibinfo{person}{Xin Nie}, \bibinfo{person}{Jiaopu Wen}, {and} \bibinfo{person}{Yangguang Yan}.} \bibinfo{year}{2013}\natexlab{}.
\newblock \showarticletitle{Overview of SiC power devices and its applications in power electronic converters}. In \bibinfo{booktitle}{\emph{2013 IEEE 8th Conference on Industrial Electronics and Applications (ICIEA)}}. \bibinfo{pages}{466--471}.
\newblock
\urldef\tempurl%
\url{https://doi.org/10.1109/ICIEA.2013.6566414}
\showDOI{\tempurl}


\bibitem[Raworth(2017)]%
        {Raworth2017ADF}
\bibfield{author}{\bibinfo{person}{Kate Raworth}.} \bibinfo{year}{2017}\natexlab{}.
\newblock \showarticletitle{A Doughnut for the Anthropocene: humanity's compass in the 21st century.}
\newblock \bibinfo{journal}{\emph{The Lancet. Planetary health}}  \bibinfo{volume}{1 2} (\bibinfo{year}{2017}), \bibinfo{pages}{e48--e49}.
\newblock
\urldef\tempurl%
\url{https://api.semanticscholar.org/CorpusID:46919938}
\showURL{%
\tempurl}


\bibitem[Rezaeizadeh and Mastellone(2024a)]%
        {aminCDC}
\bibfield{author}{\bibinfo{person}{Amin Rezaeizadeh} {and} \bibinfo{person}{Silvia Mastellone}.} \bibinfo{year}{2024}\natexlab{a}.
\newblock \showarticletitle{Reliability and Lifetime Optimal Control for Electric Vehicle Power Converters}.
\newblock \bibinfo{journal}{\emph{IEEE Control Systems Letters}}  \bibinfo{volume}{8} (\bibinfo{year}{2024}), \bibinfo{pages}{2099--2104}.
\newblock
\urldef\tempurl%
\url{https://doi.org/10.1109/LCSYS.2024.3419811}
\showDOI{\tempurl}


\bibitem[Rezaeizadeh and Mastellone(2024b)]%
        {pcim}
\bibfield{author}{\bibinfo{person}{Amin Rezaeizadeh} {and} \bibinfo{person}{Silvia Mastellone}.} \bibinfo{year}{2024}\natexlab{b}.
\newblock \showarticletitle{Reliability-Optimized Space Vector Modulation (RO-SVM) for Semiconductors Lifetime Enhancement}. In \bibinfo{booktitle}{\emph{PCIM Europe 2024; International Exhibition and Conference for Power Electronics, Intelligent Motion, Renewable Energy and Energy Management}}. \bibinfo{pages}{1681--1687}.
\newblock
\urldef\tempurl%
\url{https://doi.org/10.30420/566262230}
\showDOI{\tempurl}


\bibitem[Rezaeizadeh and Mastellone(2025)]%
        {Amin_short_note}
\bibfield{author}{\bibinfo{person}{A. Rezaeizadeh} {and} \bibinfo{person}{S. Mastellone}.} \bibinfo{year}{2025}\natexlab{}.
\newblock \bibinfo{booktitle}{\emph{A short note on frequency-domain damage calculation for power semiconductors reliability control design}}.
\newblock \bibinfo{type}{{T}echnical {R}eport}. \bibinfo{institution}{Fachhochschule Nordwestschweiz FHNW}, \bibinfo{address}{Windisch}.
\newblock
\urldef\tempurl%
\url{https://doi.org/10.26041/fhnw-11938}
\showDOI{\tempurl}


\bibitem[Rezaeizadeh et~al\mbox{.}(2024)]%
        {aminECC2024_mobility}
\bibfield{author}{\bibinfo{person}{Amin Rezaeizadeh}, \bibinfo{person}{Gioele Zardini}, \bibinfo{person}{Silvia Mastellone}, \bibinfo{person}{Andrea Censi}, {and} \bibinfo{person}{Emilio Frazzoli}.} \bibinfo{year}{2024}\natexlab{}.
\newblock \showarticletitle{Reliability-aware Control of Power Converters in Mobility Applications}. In \bibinfo{booktitle}{\emph{2024 European Control Conference (ECC)}}.
\newblock


\bibitem[Sangwongwanich and Blaabjerg(2021)]%
        {Frede2}
\bibfield{author}{\bibinfo{person}{Ariya Sangwongwanich} {and} \bibinfo{person}{Frede Blaabjerg}.} \bibinfo{year}{2021}\natexlab{}.
\newblock \showarticletitle{Monte Carlo Simulation With Incremental Damage for Reliability Assessment of Power Electronics}.
\newblock \bibinfo{journal}{\emph{IEEE Transactions on Power Electronics}} \bibinfo{volume}{36}, \bibinfo{number}{7} (\bibinfo{year}{2021}), \bibinfo{pages}{7366--7371}.
\newblock
\urldef\tempurl%
\url{https://doi.org/10.1109/TPEL.2020.3044438}
\showDOI{\tempurl}


\bibitem[Schweizer et~al\mbox{.}(2013)]%
        {abb_3L_eff}
\bibfield{author}{\bibinfo{person}{Mario Schweizer}, \bibinfo{person}{Thomas Friedli}, {and} \bibinfo{person}{Johann~W. Kolar}.} \bibinfo{year}{2013}\natexlab{}.
\newblock \showarticletitle{Comparative Evaluation of Advanced Three-Phase Three-Level Inverter/Converter Topologies Against Two-Level Systems}.
\newblock \bibinfo{journal}{\emph{IEEE Transactions on Industrial Electronics}} \bibinfo{volume}{60}, \bibinfo{number}{12} (\bibinfo{year}{2013}), \bibinfo{pages}{5515--5527}.
\newblock
\urldef\tempurl%
\url{https://doi.org/10.1109/TIE.2012.2233698}
\showDOI{\tempurl}


\bibitem[Tang et~al\mbox{.}(2021)]%
        {Tang2021ReliabilityStudy}
\bibfield{author}{\bibinfo{person}{Qian Tang}, \bibinfo{person}{Xiong Shu}, \bibinfo{person}{Guanghui Zhu}, \bibinfo{person}{Jiande Wang}, {and} \bibinfo{person}{Huan Yang}.} \bibinfo{year}{2021}\natexlab{}.
\newblock \showarticletitle{{Reliability study of bev powertrain system and its components—a case study}}.
\newblock \bibinfo{journal}{\emph{Processes}} \bibinfo{volume}{9}, \bibinfo{number}{5} (\bibinfo{year}{2021}).
\newblock
\showISSN{22279717}
\urldef\tempurl%
\url{https://doi.org/10.3390/pr9050762}
\showDOI{\tempurl}


\bibitem[van~der Broeck et~al\mbox{.}(2019)]%
        {8540418}
\bibfield{author}{\bibinfo{person}{Christoph~H. van~der Broeck}, \bibinfo{person}{Lukas~A. Ruppert}, \bibinfo{person}{Robert~D. Lorenz}, {and} \bibinfo{person}{Rik~W. De~Doncker}.} \bibinfo{year}{2019}\natexlab{}.
\newblock \showarticletitle{Methodology for Active Thermal Cycle Reduction of Power Electronic Modules}.
\newblock \bibinfo{journal}{\emph{IEEE Transactions on Power Electronics}} \bibinfo{volume}{34}, \bibinfo{number}{8} (\bibinfo{year}{2019}), \bibinfo{pages}{8213--8229}.
\newblock
\urldef\tempurl%
\url{https://doi.org/10.1109/TPEL.2018.2882184}
\showDOI{\tempurl}


\bibitem[Wang et~al\mbox{.}(2013)]%
        {6532474}
\bibfield{author}{\bibinfo{person}{Huai Wang}, \bibinfo{person}{Marco Liserre}, {and} \bibinfo{person}{Frede Blaabjerg}.} \bibinfo{year}{2013}\natexlab{}.
\newblock \showarticletitle{Toward Reliable Power Electronics: Challenges, Design Tools, and Opportunities}.
\newblock \bibinfo{journal}{\emph{IEEE Industrial Electronics Magazine}} \bibinfo{volume}{7}, \bibinfo{number}{2} (\bibinfo{year}{2013}), \bibinfo{pages}{17--26}.
\newblock
\urldef\tempurl%
\url{https://doi.org/10.1109/MIE.2013.2252958}
\showDOI{\tempurl}


\bibitem[Wang et~al\mbox{.}(2014)]%
        {bondwire1}
\bibfield{author}{\bibinfo{person}{Huai Wang}, \bibinfo{person}{Marco Liserre}, \bibinfo{person}{Frede Blaabjerg}, \bibinfo{person}{Peter de Place~Rimmen}, \bibinfo{person}{John~B. Jacobsen}, \bibinfo{person}{Thorkild Kvisgaard}, {and} \bibinfo{person}{Jørn Landkildehus}.} \bibinfo{year}{2014}\natexlab{}.
\newblock \showarticletitle{Transitioning to Physics-of-Failure as a Reliability Driver in Power Electronics}.
\newblock \bibinfo{journal}{\emph{IEEE Journal of Emerging and Selected Topics in Power Electronics}} \bibinfo{volume}{2}, \bibinfo{number}{1} (\bibinfo{year}{2014}), \bibinfo{pages}{97--114}.
\newblock
\urldef\tempurl%
\url{https://doi.org/10.1109/JESTPE.2013.2290282}
\showDOI{\tempurl}


\bibitem[Wang et~al\mbox{.}(2012)]%
        {Frede1}
\bibfield{author}{\bibinfo{person}{Huai Wang}, \bibinfo{person}{Ke Ma}, {and} \bibinfo{person}{Frede Blaabjerg}.} \bibinfo{year}{2012}\natexlab{}.
\newblock \showarticletitle{Design for reliability of power electronic systems}. In \bibinfo{booktitle}{\emph{IECON 2012 - 38th Annual Conference on IEEE Industrial Electronics Society}}. \bibinfo{pages}{33--44}.
\newblock
\urldef\tempurl%
\url{https://doi.org/10.1109/IECON.2012.6388833}
\showDOI{\tempurl}


\bibitem[Wintrich and Scheuermann(2021)]%
        {semikron}
\bibfield{author}{\bibinfo{person}{Arendt Wintrich} {and} \bibinfo{person}{Uwe Scheuermann}.} \bibinfo{year}{2021}\natexlab{}.
\newblock \bibinfo{booktitle}{\emph{Power Cycle Model for {IGBT} Product Lines}}.
\newblock \bibinfo{type}{Application Note} AN21-001. \bibinfo{institution}{Semikron Danfoss}, \bibinfo{address}{Nuremberg, Germany}.
\newblock


\bibitem[Wolfgang(2007)]%
        {failure1}
\bibfield{author}{\bibinfo{person}{E Wolfgang}.} \bibinfo{year}{2007}\natexlab{}.
\newblock \showarticletitle{Examples for failures in power electronics systems}.
\newblock \bibinfo{journal}{\emph{ECPE tutorial on reliability of power electronic systems, Nuremberg, Germany}} (\bibinfo{year}{2007}), \bibinfo{pages}{19--20}.
\newblock


\bibitem[Wu et~al\mbox{.}(1996)]%
        {bondwire3}
\bibfield{author}{\bibinfo{person}{Wuchen Wu}, \bibinfo{person}{Guo Gao}, \bibinfo{person}{Limin Dong}, \bibinfo{person}{Zhengyuan Wang}, \bibinfo{person}{M. Held}, \bibinfo{person}{P. Jacob}, {and} \bibinfo{person}{P. Scacco}.} \bibinfo{year}{1996}\natexlab{}.
\newblock \showarticletitle{Thermal reliability of power insulated gate bipolar transistor (IGBT) modules}. In \bibinfo{booktitle}{\emph{Twelfth Annual IEEE Semiconductor Thermal Measurement and Management Symposium. Proceedings}}. \bibinfo{pages}{136--141}.
\newblock
\urldef\tempurl%
\url{https://doi.org/10.1109/STHERM.1996.545103}
\showDOI{\tempurl}


\bibitem[Yang and Saeedifard(2017)]%
        {thermalControl1}
\bibfield{author}{\bibinfo{person}{Qichen Yang} {and} \bibinfo{person}{Maryam Saeedifard}.} \bibinfo{year}{2017}\natexlab{}.
\newblock \showarticletitle{Active thermal loading control of the modular multilevel converter by a multi-objective optimization method}. In \bibinfo{booktitle}{\emph{IECON 2017 - 43rd Annual Conference of the IEEE Industrial Electronics Society}}. \bibinfo{pages}{4482--4487}.
\newblock
\urldef\tempurl%
\url{https://doi.org/10.1109/IECON.2017.8216772}
\showDOI{\tempurl}


\bibitem[Yang et~al\mbox{.}(2009a)]%
        {semi_survey}
\bibfield{author}{\bibinfo{person}{Shaoyong Yang}, \bibinfo{person}{Angus Bryant}, \bibinfo{person}{Phil Mawby}, \bibinfo{person}{Dawei Xiang}, \bibinfo{person}{Li Ran}, {and} \bibinfo{person}{Peter Tavner}.} \bibinfo{year}{2009}\natexlab{a}.
\newblock \showarticletitle{An industry-based survey of reliability in power electronic converters}. In \bibinfo{booktitle}{\emph{2009 IEEE Energy Conversion Congress and Exposition}}. \bibinfo{pages}{3151--3157}.
\newblock
\urldef\tempurl%
\url{https://doi.org/10.1109/ECCE.2009.5316356}
\showDOI{\tempurl}


\bibitem[Yang et~al\mbox{.}(2009b)]%
        {failure2}
\bibfield{author}{\bibinfo{person}{Shaoyong Yang}, \bibinfo{person}{Angus Bryant}, \bibinfo{person}{Phil Mawby}, \bibinfo{person}{Dawei Xiang}, \bibinfo{person}{Li Ran}, {and} \bibinfo{person}{Peter Tavner}.} \bibinfo{year}{2009}\natexlab{b}.
\newblock \showarticletitle{An industry-based survey of reliability in power electronic converters}. In \bibinfo{booktitle}{\emph{2009 IEEE Energy Conversion Congress and Exposition}}. \bibinfo{pages}{3151--3157}.
\newblock
\urldef\tempurl%
\url{https://doi.org/10.1109/ECCE.2009.5316356}
\showDOI{\tempurl}


\end{thebibliography}


\end{document}